# A dynamic high-frequency consistent continualization of beam-lattice materials

Andrea Bacigalupo[1] and Luigi Gambarotta[1*]

[1]Department of Civil, Chemical and Environmental Engineering, University of Genova, Italy

**Abstract**

The main purpose of the present paper is to solve the thermodynamic inconsistencies that result when deriving equivalent micropolar models of periodic beam-lattice materials through standard continualization schemes. In fact, this technique identifies higher order micropolar continua characterized by non-positive defined elastic potential energy. Despite this, such models are capable of accurately simulating the optical branches of the discrete Lagrangian model, a property lacking in the thermodynamically consistent standard micropolar continuum. To overcome these thermodynamic inconsistencies while preserving good simulations of the frequency band structure, a dynamic high-frequency consistent continualization is proposed. This continualization scheme is based on a first order regularization approach coupled with a suitable transformation of the difference equation of motion of the discrete Lagrangian system into pseudo-differential equations. A formal Taylor expansion of the pseudo-differential operators allows to obtain differential field equations at various orders according to the continualization order. Thermodynamically consistent higher order micropolar continua with non-local elasticity and inertia are obtained. Finally, the convergence of the frequency band structure of the higher order micropolar models to that of the discrete Lagrangian system is shown as the continualization order increases.

________________________________________________
[*] Corresponding Author, luigi.gambarotta@unige.it

# 1. Introduction

Lattice microstructures have long inspired researchers to achieve innovative materials with efficient and exotic overall mechanical performance (see Gibson and Ashby, 1997; Fleck *et al.*, 2010), focusing on both the elastic-acoustic behavior (see Martisson and Movchan, 2003) and on the inelastic properties (see Chen *et al.*, 1998; Seiler *et al.*, 2021). By restricting the focus to the elastic properties, several contributions concern auxetic lattice microstructures (see Krödel *et al.*, 2014; Lu *et al.*, 2017), chiral lattices (see Prall and Lakes, 1997; Chen *et al.*, 2014; Bacigalupo and Gambarotta, 2016) and acoustic properties of the lattice materials and their control (see Phani *et al.*, 2006; Carta *et al.*, 2019; Piccolroaz *et al.*, 2017; Bordiga *et al.*, 2021). Due to the high number of degrees of freedom characterizing lattice materials and the need for synthetic representations of the mechanical response, standard continuum models based on homogenization techniques may be preferred to catch both static and dynamic properties of such materials (see for instance Ostoja-Starzewski, 2002). When size-effects, dispersive waves propagation and boundary layer effects have to be taken into account to simulate the discrete models, non-local equivalent continuous models have to be derived.

Homogeneous non-local continuum models equivalent to lattices with central interactions have been formulated by Askes and Metrikine (2005) among the others. These models are identified via the standard continualization of the micro-displacement field, i.e. by assuming a down scaling law in which the micro displacement field is expressed as a truncated Taylor expansion of the macro-displacement field.

When the lattice ligaments exhibit axial and bending stiffness, being described as Bernoulli-Euler beams, the nodal displacements of the resulting beam-lattice are enriched with the rotational degree of freedom together with rotational inertial effects. A complex behavior due to the coupling of translational and rotational modes is obtained and the dynamic response is characterized by acoustic and optical branches in the Floquet-Bloch spectrum. This problem has been faced by Suiker *et al.* (2001) who derived a micropolar continuum equivalent to two-dimensional periodic lattices with massless ligaments by applying a standard continualization of the generalized micro-displacement field. Gonella and Ruzzene (2008), and Lombardo and Askes (2012) addressed the case in which the nodal rotational inertia is neglected. An improved homogenization technique based on a multi-field approach has been formulated by Vasiliev *et al.* (2008, 2010) and applied to square lattices endowed of rotational inertia. An asymptotic homogenization to obtain micropolar continuum models of beam-lattice has been proposed by Dos Reis and Ganghoffer (2012) and applied to structural simulations. A further high contrast homogenization technique to obtain a high frequency approximation of the band structure of periodic lattice materials has been formulated in Kamotski and Smyshlyaev (2019).

Despite these theoretical contributions, some problems in the micropolar homogenization of beam-lattice based on continualization techniques of the discrete equations of the Lagrangian system still seem open. In fact, Bacigalupo and Gambarotta (2017) have shown that the equivalent micropolar continuum of a beam-lattice obtained through the standard continualization (see Suiker *et al.*, 2001) turns out to be thermodynamically inconsistent, i.e. the elastic energy density due to micro-curvatures turns out to be negative defined, a question already highlighted by Bazant and Christensen (1972) and Kumar and McDowell (2004). On the other hand, the analysis by Bacigalupo and Gambarotta (2017) concerning the dynamic dispersive properties of the homogeneous model has shown that this micropolar model provides good simulations of the dispersion functions of the reference discrete Lagrangian system, both for the acoustic branches and for the optical branch. A property lacking in the thermodynamically consistent standard micropolar continuum.

The need for a thermodynamically consistent micropolar continuum capable of simulating both the static and dynamic response of the discrete lattice systems motivates the search for a proper continualization technique. To this end, the enhanced continualization technique proposed by Bacigalupo and Gambarotta (2019) for one-dimensional lattice and by Bacigalupo and Gambarotta (2021) for two-dimensional lattices undergoing transversal motion is here developed. This technique, which has been also successfully tested in comparative studies by Gómez-Silva et al. (2020) and Gómez-Silva and Zaera (2021), is here applied to a representative case of square beam-lattice to obtain the field equations governing the motion of the equivalent non-local continuum formulated at different orders. Through the presented approach both non-local stiffness and inertia terms are obtained, in agreement with the non-local continuum models proposed by the seminal papers of Mindlin (1964), Eringen (1983), Askes and Aifantis (2011), and more recently by Bacigalupo and Gambarotta (2014), De Domenico and Askes (2016), and De Domenico *et al.* (2019).

The paper is organized as follows. In Section 2, the linear equations of motion of the discrete Lagrangian system representative of the square lattice made up with massless beams are formulated together with the equations governing the free undamped propagation of elastic harmonic waves. In Section 3, the standard continualization of the Lagrangian model is formulated and the linear equations of motion of the equivalent micropolar continuum are derived. The constitutive equations are derived and the Floquet-Bloch spectra obtained. Moreover, some pathologic limitations that can rise up in the static and dynamic fields are discussed. In Section 4, the enhanced continualization scheme is presented together with the downscaling law and the mathematical procedure to derive the equations of motion at increasing orders is outlined, focusing on the sources of constitutive and inertial non-localities. In Section 5, some benchmark problems concerning the acoustic properties of the lattice are stated and solved in order to discuss and highlight the potential of the equivalent non-local continuum at the different orders and its validity limits. Concluding remarks are finally pointed out.

**2. Square beam lattices: equation of motion and harmonic wave propagation**

Let consider a square beam-lattice made up of massless ligaments of thickness $w$, length $l$, Young modulus $E_s$. Each node has mass $M$ and moment of inertia $J$. Let consider a reference node and the four surrounding nodes connected to it denoted by index $i=1,..,4$, as shown in see Figure 1. The motion of the reference node and $i$-th node is represented by the time-dependent vector $\mathbf{u}(t) = \{u(t) \quad v(t) \quad \phi(t)\}^T$ and $\mathbf{u}_i(t) = \{u_i(t) \quad v_i(t) \quad \phi_i(t)\}^T$, respectively, while generalized forces applied to the reference node are represented by the force $\mathbf{f}$ and the couple $c$. The equation of motion of the reference node are derived (see for instance Gambarotta and Bacigalupo, 2017) and are written in a convenient form in terms of the components of the non-dimensional generalized displacements vector $\mathbf{v}(t) = \{\zeta(t) \quad \psi(t) \quad \phi(t)\}^T$ as follows

$$\begin{cases} \zeta_1 + \zeta_3 - 2(1+r^2)\zeta + r^2(\zeta_2 + \zeta_4) + \dfrac{r^2}{2}(\phi_2 - \phi_4) + \hat{f}_\zeta - J_m \ddot{\zeta} = 0 \\ \psi_2 + \psi_4 - 2(1+r^2)\psi + r^2(\psi_1 + \psi_3) + \dfrac{r^2}{2}(\phi_3 - \phi_1) + \hat{f}_\psi - J_m \ddot{\psi} = 0, \\ \dfrac{r^2}{2}(\psi_1 - \psi_3 - \xi_2 + \xi_4) - \dfrac{r^2}{6}(8\phi + \phi_1 + \phi_2 + \phi_3 + \phi_4) + \hat{c} - J_R \ddot{\phi} = 0 \end{cases} \quad (1)$$

being $\zeta = u/l$, $\psi = v/l$ the non-dimensional displacements, $r = w/l$ the ratio between the beam thickness and the ligament length, $J_m = M/(rE_s)$ and $J_R = J/(rE_s l^2) = J_m/\eta$ the non-dimensional node mass and moment of inertia, $\hat{f}_\zeta$, $\hat{f}_\psi$ and $\hat{c}$ the non-dimensional generalized forces applied to the reference node. Moreover, the non-dimensional parameter $\eta = (l/R)^2$ depends on the radius of gyration $R$ of the nodal mass.

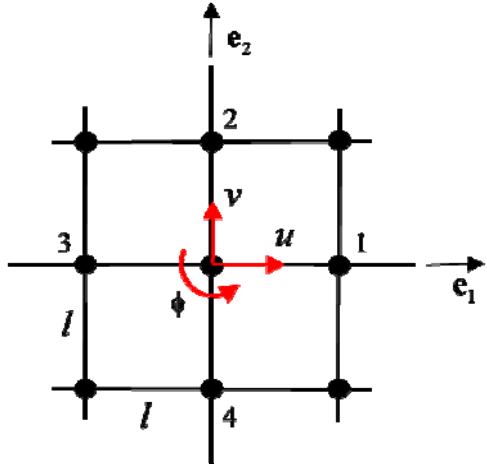

Figure 1: Square beam-lattice: nodes and displacements.

The propagation of harmonic waves in the square beam-lattice is investigated imposing the motion $\mathbf{v}(t) = \tilde{\mathbf{v}} \exp(-I\omega t)$ in the reference node and the motion $\mathbf{v}_i(\mathbf{x},t) = \tilde{\mathbf{v}} \exp[I(\mathbf{k}\cdot\mathbf{x}_i - \omega t)]$ in the $i$-th adjacent node, with polarization vector $\tilde{\mathbf{v}} = \{\tilde{\zeta} \quad \tilde{\psi} \quad \tilde{\phi}\}^T$, angular frequency $\omega$, wavevector $\mathbf{k} = \{k_1 \quad k_2\}^T$, imaginary unit $I$, and denoting with $\mathbf{x}_i = l\,\mathbf{n}_i$ the vector connecting the reference node to $i$-th adjacent one, being $\mathbf{n}_i$ the unit vector associated to the $i$-th ligament. Therefore, the free wave propagation is governed by the generalized eigenproblem

$$\left(\mathbf{K}(\mathbf{k}) - \omega^2 \mathbf{M}\right)\tilde{\mathbf{v}} = \left(\begin{bmatrix} K_{\zeta\zeta} & K_{\zeta\psi} & K_{\zeta\phi} \\ K_{\psi\zeta} & K_{\psi\psi} & K_{\psi\phi} \\ K_{\phi\zeta} & K_{\phi\psi} & K_{\phi\phi} \end{bmatrix} - \omega^2 \begin{bmatrix} J_m & 0 & 0 \\ 0 & J_m & 0 \\ 0 & 0 & \dfrac{J_m}{\eta} \end{bmatrix}\right) \begin{Bmatrix} \tilde{\zeta} \\ \tilde{\psi} \\ \tilde{\phi} \end{Bmatrix} = \mathbf{0}, \tag{2}$$

where the non-vanishing components of the wavevector-dependent Hermitian matrix $\mathbf{K}(\mathbf{k})$ are

$$\begin{aligned}
K_{\zeta\zeta}(\mathbf{k}) &= 2\left[1-\cos(k_1 l)\right] + 2r^2\left[1-\cos(k_2 l)\right], \\
K_{\zeta\phi}(\mathbf{k}) &= -K_{\phi\zeta}(\mathbf{k}) = -Ir^2 \sin(k_2 l), \\
K_{\zeta\zeta}(\mathbf{k}) &= 2\left[1-\cos(k_2 l)\right] + 2r^2\left[1-\cos(k_1 l)\right], \\
K_{\zeta\phi}(\mathbf{k}) &= -K_{\phi\zeta}(\mathbf{k}) = Ir^2 \sin(k_1 l), \\
K_{\phi\phi}(\mathbf{k}) &= \dfrac{r^2}{3}\left[4 + \cos(k_1 l) + \cos(k_1 l)\right],
\end{aligned} \tag{3}$$

whose solution is characterized by three dispersion functions $\omega_h(\mathbf{k})$, $h=1,3$, and the corresponding non-dimensional generalized polarization vector $\tilde{\mathbf{v}}_h(\mathbf{k})$. Specifically, when considering the limit of long wavelengths, namely for $\|\mathbf{k}\| \to 0$, two dispersion functions, corresponding to the acoustic spectral branches, tend to zero $\omega_{ac1,2}(\|\mathbf{k}\| \to 0) = 0$, whereas the third dispersion function, corresponding to the optical spectral branch, tends to a critical point $\omega_{opt}(\|\mathbf{k}\| \to 0) = \sqrt{\eta K_{\phi\phi}(\mathbf{k}=\mathbf{0})/J_m} = \sqrt{2\eta r^2/J_m}$ with vanishing group velocity.

## 3. Micropolar homogenization from the standard continualization approach

A continuum model equivalent to the discrete Lagrangian system presented in Section 2 is here derived through a continualization approach according to the seminal paper by Bazant and Christensen (1972). The non-dimensional time-dependent displacements and the rotation $\phi_i(t)$ of the $i$-th node are collected in vector $\mathbf{v}_i(t) = \{\zeta_i(t) \quad \psi_i(t) \quad \phi_i(t)\}^T$ approximated through a second order expansion of the non-dimensional generalized macro-displacement fields $\Upsilon(\mathbf{x},t) = \{Z(\mathbf{x},t) \quad \Psi(\mathbf{x},t) \quad \Phi(\mathbf{x},t)\}^T$ of the continuum as follows

$$\mathbf{v}_i(t) \cong \Upsilon(\mathbf{x},t) + l\, \mathbf{H}(\mathbf{x},t)\mathbf{n}_i + \frac{1}{2}l^2\, \nabla \mathbf{H}(\mathbf{x},t):(\mathbf{n}_i \otimes \mathbf{n}_i) \quad , \tag{4}$$

$\mathbf{H} = \nabla \Upsilon$ and $\nabla \mathbf{H}$ being the generalized macro-displacement gradient and its second gradient, respectively. By substituting the approximations (4) in equation (1), the equation of motion of a micropolar equivalent continuum are obtained in the form

$$\begin{cases} l^2 \dfrac{\partial^2 Z}{\partial x_1^2} + r^2 l \dfrac{\partial}{\partial x_2}\left(l\dfrac{\partial Z}{\partial x_2} + \Phi\right) + \hat{f}_\xi - J_m \ddot{\Xi} = 0 \\ l^2 \dfrac{\partial^2 \Psi}{\partial x_2^2} + r^2 l \dfrac{\partial}{\partial x_1}\left(l\dfrac{\partial \Psi}{\partial x_1} - \Phi\right) + \hat{f}_\psi - J_m \ddot{\Psi} = 0 \\ -\dfrac{1}{6}r^2 l^2 \Delta\Phi + r^2\left(l\dfrac{\partial \Psi}{\partial x_1} - \Phi\right) - r^2\left(l\dfrac{\partial Z}{\partial x_2} + \Phi\right) + \hat{c} - \dfrac{J_m}{\eta}\ddot{\Phi} = 0 \end{cases} \tag{5}$$

It is worth to note, in agreement with Bazant and Christensen (1972) and Kumar and McDowell (2004), that the elastic potential energy density of the micropolar continuum identified via standard continualization is non positive defined. Specifically, the contribution associated to the third equation of PDE system (5) is non positive defined due to the negative sign of the term involving the Laplacian operator (see also Bacigalupo and Gambarotta, 2017).

The harmonic wave propagation is analysed by imposing the solution form $\Upsilon(\mathbf{x},t) = \tilde{\Upsilon} \exp[I(\mathbf{k} \cdot \mathbf{x} - \omega t)]$, being $\tilde{\Upsilon} = \{\tilde{Z} \quad \tilde{\Psi} \quad \tilde{\Phi}\}^T$ the polarization vector, in the governing equation (5) and deriving in this way the following eigenproblem in terms of the non-dimensional generalized macro-displacements vector $\Upsilon(\mathbf{x},\tau) = \{Z(\mathbf{x},\tau) \quad \Psi(\mathbf{x},\tau) \quad \Phi(\mathbf{x},\tau)\}^T$ as follows

$$\left(\mathbf{L}(\mathbf{k}) - \omega^2 \mathbf{M}\right)\tilde{\Upsilon} = \left(\begin{bmatrix} L_{ZZ} & L_{Z\Psi} & L_{Z\Phi} \\ L_{\Psi Z} & L_{\Psi\Psi} & L_{\Psi\Phi} \\ L_{\Phi Z} & L_{\Phi\Psi} & L_{\Phi\Phi} \end{bmatrix} - \omega^2 \begin{bmatrix} J_m & 0 & 0 \\ 0 & J_m & 0 \\ 0 & 0 & \dfrac{J_m}{\eta} \end{bmatrix}\right) \begin{Bmatrix} \tilde{Z} \\ \tilde{\Psi} \\ \tilde{\Phi} \end{Bmatrix} = \mathbf{0}, \quad (6)$$

where the non-vanishing components of the wavevector-dependent Hermitian matrix $\mathbf{L}(\mathbf{k})$ are

$$\begin{aligned}
L_{ZZ}(\mathbf{k}) &= k_1^2 l^2 + r^2 k_2^2 l^2, \\
L_{Z\Phi}(\mathbf{k}) &= -L_{\Phi Z}(\mathbf{k}) = -Ir^2 k_2 l, \\
L_{\Psi\Psi}(\mathbf{k}) &= k_2^2 l^2 + r^2 k_1^2 l^2, \\
L_{\Psi\Phi}(\mathbf{k}) &= -L_{\Phi\Psi}(\mathbf{k}) = Ir^2 k_1 l, \\
L_{\Phi\Phi}(\mathbf{k}) &= \frac{r^2}{3}\left[-\frac{1}{2}(k_1^2 l^2 + k_2^2 l^2) + 6\right].
\end{aligned} \quad (7)$$

It is easy to recognize that the matrix $\mathbf{L}(\mathbf{k})$ is the second order series expansions in the wavevector components $k_1$, $k_2$ of the matrix $\mathbf{K}(\mathbf{k})$ governing the frequency band structures of the discrete Lagrangian system (see for details Bacigalupo and Gambarotta, 2017). As minor remark, when considering the limit of long wavelengths, namely for $\|\mathbf{k}\| \to 0$, the acoustic spectral branches tend to zero $\omega_{ac1,2}(\|\mathbf{k}\| \to 0) = 0$, whereas the third dispersion function, corresponding to the optical spectral branch, tends to a critical point $\omega_{opt}(\|\mathbf{k}\| \to 0) = \sqrt{\eta L_{\Phi\Phi}(\mathbf{k}=\mathbf{0})/J_m}$ with vanishing group velocity. It can be observed that the optical critical point $\omega_{opt}(\|\mathbf{k}\| \to 0)$ of the discrete and continuum models coincide to each other, since the equality

$L_{\Phi\Phi}(\mathbf{k}=\mathbf{0}) = K_{\phi\phi}(\mathbf{k}=\mathbf{0})$ holds.

From the equation of motion (5) the dimensional constitutive equation of the homogenized micropolar continuum may be easily derived in terms of strain and curvature components and corresponding stresses and micro-couples as follows

$$\begin{Bmatrix} \sigma_{11} \\ \sigma_{22} \\ \sigma_{12} \\ \sigma_{21} \\ m_1 \\ m_2 \end{Bmatrix} = \begin{bmatrix} 2\mu & 0 & 0 & 0 & 0 & 0 \\ 0 & 2\mu & 0 & 0 & 0 & 0 \\ 0 & 0 & \kappa & 0 & 0 & 0 \\ 0 & 0 & 0 & \kappa & 0 & 0 \\ 0 & 0 & 0 & 0 & S & 0 \\ 0 & 0 & 0 & 0 & 0 & S \end{bmatrix} \begin{Bmatrix} \gamma_{11} \\ \gamma_{22} \\ \gamma_{12} \\ \gamma_{21} \\ \chi_1 \\ \chi_2 \end{Bmatrix},$$

(8)

being $\mu = E_s r/2$, $\kappa = E_s r^3$, $S = -E_s r^3 l^2/6$ the overall elastic moduli, the last modulus being negative independently on the lattice parameters.

To obtain positive defined energy density of the equivalent medium, the constitutive equation has been derived by several Authors through an application of the generalized Hill-Mandel lemma. In this case the micropolar elastic constant turns out to be $S^+ = E_s r^3 l^2/3$. However, it has been shown by Bazant and Christensen (1972) and Bacigalupo and Gambarotta (2017) the latter in a more general way, that the derivation of the constitutive model through an extended Hamiltonian approach, which is equivalent to a rigorous application of the Hill-Mandell procedure, provides the same equivalent constitutive model (8) and governing equations (5) from the classical continualization approach. Moreover, if the energetically consistent formulation based on the positive micropolar elastic modulus $S^+ = E_s r^3 l^2/3$ is assumed, the optical dispersion function provided by the micropolar continuum does not qualitative agrees with the corresponding of the discrete Lagrangian one. In fact, while the component $L_{\Phi\Phi}(\mathbf{k}) = 2r^2 - \frac{1}{6} r^2 \left( k_1^2 l^2 + k_2^2 l^2 \right) = 2r^2 + S\left( k_1^2 + k_2^2 \right)/(rE_s)$ is decreasing when departing from the long wavelength condition, namely for $\|\mathbf{k}\|$ increasing, conversely an increasing trend is obtained when is assuming the positive constitutive parameter $S^+$, namely an increase of the optical dispersive function as commonly shown in literature (see for details Bacigalupo and Gambarotta, 2017).

This analysis highlights a contradictory situation. On the one hand, the micropolar continuum with negative defined elastic energy density more faithfully reproduces the optical branch of the reference discrete

Lagrangian model; on the other hand it results in an energetically non-consistent model that cannot be applied in static conditions. This inconsistency leads to a more in-depth investigation of the constitutive aspects of the micropolar model, which concern both the elastic and the inertial modelling, a question already addressed by other authors (see De Domenico *et al.*, 2019).

For the sake of clarity, the frequency band structures given by the discrete Lagrangian model is compared with those obtained by the homogenized models characterized by micropolar constitutive constant $S$ and $S^+$, respectively (see Figure 2). Specifically, the dispersive functions are represented both along the boundary of the non-dimensional irreducible first Brillouin zone identified by closed polygonal curve $\Gamma$ and along the boundary of an its sub-domain identified by closed polygonal curve $\Gamma^S$, for completeness. The vertices of the polygonal curve $\Gamma$ are identify by the values $\Xi_j$, $j=0,1,2$, of the arc-length $\Xi$ in the dimensionless plane $(k_1 l, k_2 l)$ (see Figure 2.c) while the vertices of $\Gamma^S$ are identify by the values $\Xi_j^S$, $j=0,1,2$, of the arc-length $\Xi^S$ (see Figure 2.d). A compact spectral description is given in terms of the non-dimensional frequency $\omega\sqrt{J_m}$ and the non-dimensional parameters $r$, $\eta$. The frequency band structures for square beam-lattice characterized by $r=3/50$ and $\eta=50$ are shown in Figures 2.a along the closed polygonal curve $\Gamma$, both the discrete Lagrangian model (black line) and the homogenized models with micropolar constitutive constant $S$ (cyan line) and $S^+$ (violet line), respectively. It is important to note that the micropolar model obtained by an extended Hamiltonian approach, and characterized by the micropolar constitutive constant $S^+$, presents a very reduced accuracy to representing the actual optical branch. A greater accuracy of this branch is provided by the micropolar model determined through the continualization procedure, and characterized by the micropolar constant $S$. Both homogenized models, on the other hand, provide good accuracy to representing the actual acoustic branches for sufficiently small values of $\|\mathbf{k}\|_2 l$, i.e. for long wavelength regime. As expected, the approximations of the acoustic branches provided by the two homogenized models tend to lose accuracy for short wavelengths, and the micropolar model obtained through continualization (cyan line) present also prevailing negative group velocity together with short-wave instability and unlimited group velocity. These circumstances imply the failure to satisfy of the Legendre–Hadamard ellipticity conditions (semi-ellipticity) in all first Brillouin zone and therefore the loss of hyperbolicity of the homogenized equation of motion. For a more complete perception of the accuracy of the two micropolar models a comparison between their frequency band structures with that of the discrete

Lagrangian model is given in Figure 2.b along the closed polygonal curve $\Gamma^S$ for the case of $r = 3/50$ and $\eta = 50$.

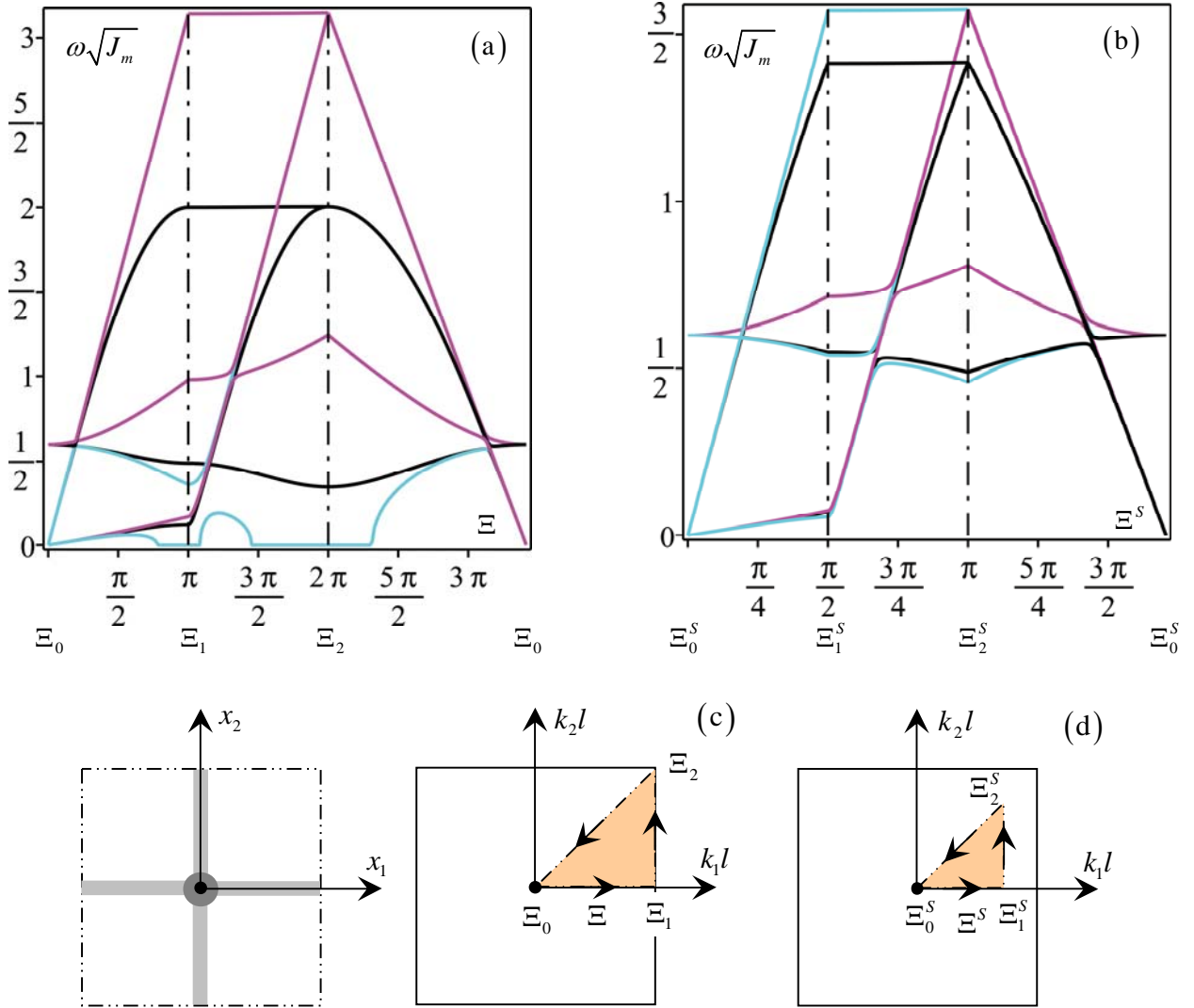

Figure 2: Frequency band structures for square beam-lattice ($r = 3/50$, $\eta = 50$). Comparison between the discrete Lagrangian model (black line) and the homogenized models characterized by micropolar constitutive constant $S$ (cyan line) and $S^+$ (violet line), respectively. (a) Dispersive functions along the boundary of the non-dimensional irreducible first Brillouin zone; (b) Dispersive functions along the boundary of the sub-domain of the non-dimensional irreducible first Brillouin zone; (c) Periodic cell and non-dimensional first Brillouin zone (highlighted in light orange the non-dimensional irreducible first Brillouin zone); (d) Sub-domain of the non-dimensional irreducible first Brillouin zone.

## 4. Non-local homogenization from the enhanced continualization approach

The equation of motion (1) of the discrete Lagrangian model is transformed by introducing the shift operator $E_i$ relating the generalized displacement vector $\mathbf{v}_i$ of the $i$-th node to the displacement vector $\mathbf{v}$ of the reference one, i.e. $\mathbf{v}_i = E_i \mathbf{v}$ (see for details Andrianov and Awrejcewicz, 2008; Bacigalupo and Gambarotta, 2019). The shift operating on node 1 is represented in the form $E_1 = \sum_{h=0}^{\infty} \frac{\ell^h}{h!} D_1^h = \exp(\ell D_1)$, with $D_1^h = \frac{\partial^h}{\partial x_1^h}$ and hence $\mathbf{v}_1 = \exp(\ell D_1)\mathbf{v}$. This formalism implies $\mathbf{v}_2 = \exp(\ell D_2)\mathbf{v}$, $\mathbf{v}_3 = \exp(-\ell D_1)\mathbf{v}$ and $\mathbf{v}_4 = \exp(-\ell D_2)\mathbf{v}$. Accordingly, the equation of motion of the discrete Lagrangian system for vanishing nodal forces and couples are rewritten as follows

$$\begin{cases} \left[\exp(\ell D_1) - 2(1+r^2) + \exp(-\ell D_1) + r^2 \exp(\ell D_2) + r^2 \exp(-\ell D_2)\right]\zeta + \\ \quad + \frac{r^2}{2}\left[\exp(\ell D_2) - \exp(-\ell D_2)\right]\phi - J_m \ddot{\zeta} = 0 \\ \left[\exp(\ell D_2) - 2(1+r^2) + \exp(-\ell D_2) + r^2 \exp(\ell D_1) + r^2 \exp(-\ell D_1)\right]\psi + \\ \quad + \frac{r^2}{2}\left[\exp(-\ell D_1) - \exp(\ell D_1)\right]\phi - J_m \ddot{\psi} = 0 \\ \frac{r^2}{2}\left[-\exp(\ell D_2) + \exp(-\ell D_2)\right]\zeta + \frac{r^2}{2}\left[\exp(\ell D_1) - \exp(-\ell D_1)\right]\psi + \\ \quad - \frac{r^2}{6}\left[8 + \exp(\ell D_1) + \exp(\ell D_2) + \exp(-\ell D_1) + \exp(-\ell D_2)\right]\phi - \frac{J_m}{\eta}\ddot{\phi} = 0 \end{cases} \quad (9)$$

In order to obtain an equivalent continuum model, the macro-displacement field $\Upsilon(\mathbf{x},t) = \{Z(\mathbf{x},t) \quad \Psi(\mathbf{x},t) \quad \Phi(\mathbf{x},t)\}^T$ must be related to the nodal displacements through a down-scaling law. Here a proper two-dimensional extension of the central difference is considered

$$\frac{\partial^2 \Upsilon}{\partial x_1 \partial x_2} = D_1 D_2 \Upsilon = \frac{\left[\exp(D_1 l) - \exp(-D_1 l)\right]\left[\exp(D_2 l) - \exp(-D_2 l)\right]}{4l^2}\mathbf{v}, \quad (10)$$

according to a general formulation proposed by Bacigalupo and Gambarotta (2021). Hence, the local displacement field may be expressed in terms of the macro-displacement field through a pseudo-differential operator

$$\upsilon = \frac{4l^2 D_1 D_2}{\left[\exp(D_1 l)-\exp(-D_1 l)\right]\left[\exp(D_2 l)-\exp(-D_2 l)\right]}\Upsilon = F(D_1,D_2)\Upsilon. \qquad (11)$$

After substituting equation (11) in (9), the equation of motion may be written as

$$\begin{bmatrix} P_{\zeta\zeta}(D_1,D_2) & 0 & P_{\zeta\phi}(D_1,D_2) \\ 0 & P_{\psi\psi}(D_1,D_2) & P_{\psi\phi}(D_1,D_2) \\ P_{\phi\zeta}(D_1,D_2) & P_{\phi\psi}(D_1,D_2) & P_{\phi\phi}(D_1,D_2) \end{bmatrix}\Upsilon - F(D_1,D_2)\begin{bmatrix} J_m & 0 & 0 \\ 0 & J_m & 0 \\ 0 & 0 & \dfrac{J_m}{\eta} \end{bmatrix}\ddot{\Upsilon} = \mathbf{0}, \qquad (12)$$

involving the following pseudo-differential operators

$$P_{\zeta\zeta}(D_1,D_2) = \left[\exp(\ell D_1) - 2(1+r^2) + \exp(-\ell D_1) + r^2\exp(\ell D_2) + r^2\exp(-\ell D_2)\right]F(D_1,D_2),$$

$$P_{\zeta\phi}(D_1,D_2) = -P_{\phi\zeta}(D_1,D_2) = \frac{r^2}{2}\left[\exp(\ell D_2) - \exp(-\ell D_2)\right]F(D_1,D_2),$$

$$P_{\psi\psi}(D_1,D_2) = \left[\exp(\ell D_2) - 2(1+r^2) + \exp(-\ell D_2) + r^2\exp(\ell D_1) + r^2\exp(-\ell D_1)\right]F(D_1,D_2), \qquad (13)$$

$$P_{\psi\phi}(D_1,D_2) = -P_{\phi\psi}(D_1,D_2) = -\frac{r^2}{2}\left[\exp(\ell D_1) - \exp(-\ell D_1)\right]F(D_1,D_2),$$

$$P_{\phi\phi}(D_1,D_2) = -\frac{r^2}{6}\left[8 + \exp(\ell D_1) + \exp(\ell D_2) + \exp(-\ell D_1) + \exp(-\ell D_2)\right]F(D_1,D_2),$$

whose Taylor series expansion in the geometric parameter $l$, that plays the role of scaling variable, are

$$F(D_1,D_2) = 1 - \frac{1}{6}l^2\frac{\partial^2}{\partial x_1^2} - \frac{1}{6}l^2\frac{\partial^2}{\partial x_2^2} + \frac{7}{360}l^4\frac{\partial^4}{\partial x_1^4} + \frac{1}{36}l^4\frac{\partial^4}{\partial x_1^2 \partial x_2^2} + \frac{7}{360}l^4\frac{\partial^4}{\partial x_2^4} + O(l^8),$$

$$P_{\zeta\zeta}(D_1,D_2) = r^2 l^2\frac{\partial^2}{\partial x_2^2} + l^2\frac{\partial^2}{\partial x_1^2} - \frac{1}{12}l^4\frac{\partial^4}{\partial x_1^4} - \frac{1}{6}l^4(r^2+1)\frac{\partial^4}{\partial x_1^2 \partial x_2^2} - \frac{1}{12}r^2 l^4\frac{\partial^4}{\partial x_2^4} + O(l^8),$$

$$P_{\zeta\phi}(D_1,D_2) = -P_{\phi\zeta}(D_1,D_2) = r^2 l\frac{\partial}{\partial x_2} - \frac{1}{6}r^2 l^3\frac{\partial^3}{\partial x_1^2 \partial x_2} + \frac{7}{360}r^2 l^5\frac{\partial^5}{\partial x_1^4 \partial x_2} + O(l^8), \qquad (14)$$

$$P_{\psi\psi}(D_1,D_2) = r^2 l^2\frac{\partial^2}{\partial x_1^2} + l^2\frac{\partial^2}{\partial x_2^2} - \frac{1}{12}r^2 l^4\frac{\partial^4}{\partial x_1^4} - \frac{1}{6}l^4(r^2+1)\frac{\partial^4}{\partial x_1^2 \partial x_2^2} - \frac{1}{12}l^4\frac{\partial^4}{\partial x_2^4} + O(l^8),$$

$$P_{\psi\phi}(D_1,D_2) = -P_{\phi\psi}(D_1,D_2) = -r^2 l\frac{\partial}{\partial x_1} + \frac{1}{6}r^2 l^3\frac{\partial^3}{\partial x_1 \partial x_2^2} - \frac{7}{360}r^2 l^5\frac{\partial^5}{\partial x_1 \partial x_2^4} + O(l^8),$$

$$P_{\phi\phi}(D_1,D_2) = -2r^2 + \frac{r^2}{6}l^2\frac{\partial^2}{\partial x_1^2} + \frac{r^2}{6}l^2\frac{\partial^2}{\partial x_2^2} - \frac{1}{40}r^2 l^4\frac{\partial^4}{\partial x_1^4} - \frac{1}{40}r^2 l^4\frac{\partial^4}{\partial x_2^4} + O(l^8).$$

If the terms of the expansion are retained up to the second order in $l$, the equation of motion takes the form

$$\begin{cases} l^2 \dfrac{\partial^2 Z}{\partial x_1^2} + r^2 l \dfrac{\partial}{\partial x_2}\left( l\dfrac{\partial Z}{\partial x_2} + \Phi \right) - J_m \ddot{Z} + \dfrac{1}{6}l^2 J_m \Delta \ddot{Z} = 0 \\ l^2 \dfrac{\partial^2 \Psi}{\partial x_2^2} + r^2 l \dfrac{\partial}{\partial x_1}\left( l\dfrac{\partial \Psi}{\partial x_1} - \Phi \right) - J_m \ddot{\Psi} + \dfrac{1}{6}l^2 J_m \Delta \ddot{\Psi} = 0 \\ \dfrac{1}{6}r^2 l^2 \Delta \Phi + r^2 \left( l\dfrac{\partial \Psi}{\partial x_1} - \Phi \right) - r^2 \left( l\dfrac{\partial Z}{\partial x_2} + \Phi \right) - \dfrac{J_m}{\eta}\ddot{\Phi} + \dfrac{1}{6}l^2 \dfrac{J_m}{\eta}\Delta \ddot{\Phi} = 0 \end{cases} \qquad (15)$$

The resulting equation of motion (15) differs from equation (5) in the positive sign of the coefficient of the Laplacian of macro-rotation and for the presence of non-local inertia terms. It easy to also to show that the elastic energy density $\Pi$ of the micropolar continuum identified via enhanced continualization takes the positive defined form

$$\Pi = \dfrac{1}{2}\left[ l^2\left(\dfrac{\partial Z}{\partial x_1}\right)^2 + l^2\left(\dfrac{\partial \Psi}{\partial x_2}\right)^2 + r^2\left( l\dfrac{\partial Z}{\partial x_2} + \Phi \right)^2 + r^2\left( l\dfrac{\partial \Psi}{\partial x_1} - \Phi \right)^2 + \dfrac{1}{6}r^2 l^2\left(\dfrac{\partial \Phi}{\partial x_1}\right)^2 + \dfrac{1}{6}r^2 l^2\left(\dfrac{\partial \Phi}{\partial x_2}\right)^2 \right]. \qquad (16)$$

Moreover, its kinetic energy density is expressed as

$$T = \dfrac{1}{2}J_m \left[ \begin{array}{l} \dot{Z}^2 + \dfrac{1}{6}l^2\left(\dfrac{\partial \dot{Z}}{\partial x_1}\right)^2 + \dfrac{1}{6}l^2\left(\dfrac{\partial \dot{Z}}{\partial x_2}\right)^2 + \dot{\Psi}^2 + \dfrac{1}{6}l^2\left(\dfrac{\partial \dot{\Psi}}{\partial x_1}\right)^2 + \dfrac{1}{6}l^2\left(\dfrac{\partial \dot{\Psi}}{\partial x_2}\right)^2 + \\ + \dfrac{1}{\eta}I_R \dot{\Phi}^2 + \dfrac{1}{6}l^2 \dfrac{1}{\eta}\left(\dfrac{\partial \dot{\Phi}}{\partial x_1}\right)^2 + \dfrac{1}{6}l^2 \dfrac{1}{\eta}\left(\dfrac{\partial \dot{\Phi}}{\partial x_2}\right)^2 \end{array} \right]. \qquad (17)$$

in terms of the macro-velocities and its first gradients and results to be positive defined. The constitutive equations are coincident with equation (8) with the exception that the positive defined micropolar constant is obtained $S_e = E_s r^3 l^2 / 6$.

The eigenproblem governing the free propagation of harmonic waves, expressed in terms of the non-dimensional generalized displacements vector $\Upsilon(\mathbf{x},\tau) = \{Z(\mathbf{x},\tau) \ \Psi(\mathbf{x},\tau) \ \Phi(\mathbf{x},\tau)\}^T$, is specialized as follows

$$\left( \mathbf{L}_e(\mathbf{k}) - \omega^2 \mathbf{M}_e(\mathbf{k}) \right)\tilde{\Upsilon} = \left( \begin{bmatrix} L^e_{ZZ} & L^e_{Z\Psi} & L^e_{Z\Phi} \\ L^e_{\Psi Z} & L^e_{\Psi\Psi} & L^e_{\Psi\Phi} \\ L^e_{\Phi Z} & L^e_{\Phi\Psi} & L^e_{\Phi\Phi} \end{bmatrix} - \omega^2 \begin{bmatrix} M^e_{ZZ} & 0 & 0 \\ 0 & M^e_{\Psi\Psi} & 0 \\ 0 & 0 & M^e_{\Phi\Phi} \end{bmatrix} \right) \begin{Bmatrix} \tilde{Z} \\ \tilde{\Psi} \\ \tilde{\Phi} \end{Bmatrix} = \mathbf{0}, \qquad (18)$$

where it is worth to note that the $\mathbf{k}$-dependent matrices $\mathbf{L}_e(\mathbf{k})$ and $\mathbf{M}_e(\mathbf{k})$ are involving non-local terms.

The non-vanishing components of $\mathbf{L}_e(\mathbf{k})$ and $\mathbf{M}_e(\mathbf{k})$ take the form

$$
\begin{aligned}
&L^e_{ZZ}(\mathbf{k}) = k_1^2 l^2 + r^2 k_2^2 l^2, \\
&L^e_{Z\Phi}(\mathbf{k}) = -L^e_{\Phi Z}(\mathbf{k}) = -Ir^2 k_2 l, \\
&L^e_{\Psi\Psi}(\mathbf{k}) = k_2^2 l^2 + r^2 k_1^2 l^2, \\
&L^e_{\Psi\Phi}(\mathbf{k}) = -L^e_{\Phi\Psi}(\mathbf{k}) = Ir^2 k_1 l, \\
&L^e_{\Phi\Phi}(\mathbf{k}) = \frac{1}{3} r^2 \left[ \frac{1}{2}\left(k_1^2 l^2 + k_2^2 l^2\right) + 6 \right], \\
&M^e_{ZZ}(\mathbf{k}) = M^e_{\Psi\Psi}(\mathbf{k}) = J_m \left[ \frac{1}{6}\left(k_1^2 l^2 + k_2^2 l^2\right) + 1 \right], \\
&M^e_{\Phi\Phi}(\mathbf{k}) = \frac{J_m}{\eta} \left[ \frac{1}{6}\left(k_1^2 l^2 + k_2^2 l^2\right) + 1 \right].
\end{aligned} \quad (19)
$$

If the terms of the expansion are retained up to the fourth order in $l$, the equation of motion takes the form

$$
\begin{cases}
+l^2 \dfrac{\partial^2 Z}{\partial x_1^2} + r^2 l \dfrac{\partial}{\partial x_2}\left(l \dfrac{\partial Z}{\partial x_2} + \Phi\right) - \dfrac{1}{6} r^2 l^3 \dfrac{\partial^3}{\partial x_1^2 \partial x_2}\left(l \dfrac{\partial Z}{\partial x_2} + \Phi\right) - \dfrac{1}{12} l^4 \dfrac{\partial^4 Z}{\partial x_1^4} - \dfrac{1}{6} l^4 \dfrac{\partial^4 Z}{\partial x_1^2 \partial x_2^2} - \dfrac{1}{12} r^2 l^4 \dfrac{\partial^4 Z}{\partial x_2^4} + \\
\quad -J_m \ddot{Z} + \dfrac{1}{6} J_m l^2 \Delta \ddot{Z} - \dfrac{7}{360} J_m l^4 \dfrac{\partial^4 \ddot{Z}}{\partial x_1^4} - \dfrac{1}{36} J_m l^4 \dfrac{\partial^4 \ddot{Z}}{\partial x_1^2 \partial x_2^2} - \dfrac{7}{360} J_m l^4 \dfrac{\partial^4 \ddot{Z}}{\partial x_2^4} = 0 \\
l^2 \dfrac{\partial^2 \Psi}{\partial x_2^2} + r^2 l \dfrac{\partial}{\partial x_1}\left(l \dfrac{\partial \Psi}{\partial x_1} - \Phi\right) - \dfrac{1}{6} r^2 l^3 \dfrac{\partial^3}{\partial x_1 \partial x_2^2}\left(l \dfrac{\partial \Psi}{\partial x_1} - \Phi\right) - \dfrac{1}{12} r^2 l^4 \dfrac{\partial^4 \Psi}{\partial x_1^4} - \dfrac{1}{6} l^4 \dfrac{\partial^4 \Psi}{\partial x_1^2 \partial x_2^2} - \dfrac{1}{12} l^4 \dfrac{\partial^4 \Psi}{\partial x_2^4} + \\
\quad -J_m \ddot{\Psi} + \dfrac{1}{6} l^2 J_m \Delta \ddot{\Psi} - \dfrac{7}{360} J_m l^4 \dfrac{\partial^4 \ddot{\Psi}}{\partial x_1^4} - \dfrac{1}{36} J_m l^4 \dfrac{\partial^4 \ddot{\Psi}}{\partial x_1^2 \partial x_2^2} - \dfrac{7}{360} J_m l^4 \dfrac{\partial^4 \ddot{\Psi}}{\partial x_2^4} = 0 \\
\dfrac{1}{6} r^2 l^2 \Delta \Phi + r^2 \left(l \dfrac{\partial \Psi}{\partial x_1} - \Phi\right) - r^2 \left(l \dfrac{\partial Z}{\partial x_2} + \Phi\right) + \dfrac{1}{6} r^2 l^3 \dfrac{\partial^3 Z}{\partial x_1^2 \partial x_2} - \dfrac{1}{6} r^2 l^3 \dfrac{\partial^3 \Psi}{\partial x_1 \partial x_2^2} - \dfrac{r^2 l^4}{40}\left(\dfrac{\partial^4 \Phi}{\partial x_1^4} + \dfrac{\partial^4 \Phi}{\partial x_2^4}\right) + \\
\quad -\dfrac{J_m}{\eta} \ddot{\Phi} + \dfrac{1}{6} \dfrac{J_m}{\eta} l^2 \Delta \ddot{\Phi} - \dfrac{7}{360} \dfrac{J_m}{\eta} l^4 \dfrac{\partial^4 \ddot{\Phi}}{\partial x_1^4} - \dfrac{1}{36} \dfrac{J_m}{\eta} l^4 \dfrac{\partial^4 \ddot{\Phi}}{\partial x_1^2 \partial x_2^2} - \dfrac{7}{360} \dfrac{J_m}{\eta} l^4 \dfrac{\partial^4 \ddot{\Phi}}{\partial x_2^4} = 0
\end{cases} \quad (20)
$$

and the non-null components of the $\mathbf{k}$-dependent matrices $\mathbf{L}_e(\mathbf{k})$ and $\mathbf{M}_e(\mathbf{k})$, that define the eigenproblem governing the free propagation of harmonic waves in the energetically consistence non-local continuum identified by 4$^{\text{th}}$ order enhanced continualization, are

$$L_{ZZ}^{e}(\mathbf{k}) = \frac{1}{12}\left[k_1^4 l^4 + r^2 k_2^4 l^4 + 2(1+r^2)k_1^2 k_2^2 l^4\right] + k_1^2 l^2 + r^2 k_2^2 l^2,$$

$$L_{Z\Phi}^{e}(\mathbf{k}) = -L_{\Phi Z}^{e}(\mathbf{k}) = -\frac{1}{6} I r^2 k_2 k_1^2 l^3 - I r^2 k_2 l,$$

$$L_{\Psi\Psi}^{e}(\mathbf{k}) = \frac{1}{12}\left[r^2 k_1^4 l^4 + k_2^4 l^4 + 2(1+r^2)k_1^2 k_2^2 l^4\right] + k_2^2 l^2 + r^2 k_1^2 l^2,$$

$$L_{\Psi\Phi}^{e}(\mathbf{k}) = -L_{\Phi\Psi}^{e}(\mathbf{k}) = \frac{1}{6} I r^2 k_1 k_2^2 l^3 + I r^2 k_1 l, \qquad (21)$$

$$L_{\Phi\Phi}^{e}(\mathbf{k}) = \frac{1}{40} r^2 \left(k_1^4 l^4 + k_2^4 l^4\right) + \frac{1}{3} r^2 \left[6 + \frac{1}{2}\left(k_1^2 l^2 + k_2^2 l^2\right)\right],$$

$$M_{ZZ}^{e}(\mathbf{k}) = M_{\Psi\Psi}^{e}(\mathbf{k}) = J_m\left(\frac{1}{360}\left[7\left(k_1^4 l^4 + k_2^4 l^4\right) + 10 k_1^2 k_2^2 l^4\right] + \frac{1}{6}\left(k_1^2 l^2 + k_2^2 l^2\right) + 1\right),$$

$$M_{\Phi\Phi}^{e}(\mathbf{k}) = \frac{J_m}{\eta}\left(\frac{1}{360}\left[7\left(k_1^4 l^4 + k_2^4 l^4\right) + 10 k_1^2 k_2^2 l^4\right] + \frac{1}{6}\left(k_1^2 l^2 + k_2^2 l^2\right) + 1\right).$$

## 6. Benchmark test for the enhanced continualization scheme

In order to show the reliability and the validity limits of the proposed enhanced continualization approach the frequency band structures of the homogenized models, for different order truncation in the characteristic length $l$ of the pseudo-differential operators involved in the integral-differential equation (12), are compared with the actual one obtained by the discrete Lagrangian model. Specifically, the non-dimensional angular frequency $\omega\sqrt{J_m}$ is expressed as function of the arch length $\Xi$ (or $\Xi^S$) measured on the closed polygonal curve $\Gamma$ (or $\Gamma^S$) with vertices identified by the values $\Xi_j$ (or $\Xi_j^S$), $j=0,1,2$, encompassing the non-dimensional first irreducible Brillouin zone (or sub-domain of non-dimensional first irreducible Brillouin zone) for different values of the non-dimensional parameters $r$, $\eta$. As the dimensionless parameter $\eta$ varies, qualitatively different frequency spectra are identified and analyzed three limit situations in which: i) crossing phenomena between the optical branch and both acoustic branches are detected (see Figure 3.a); ii) a quadratic point degeneracy, i.e. a band touching, between the optical branch and the both acoustic branches is identified (see Figure 4.a); iii) low frequency band gap between the second acoustic branch and the optical branch is obtained (see Figure 5.a).

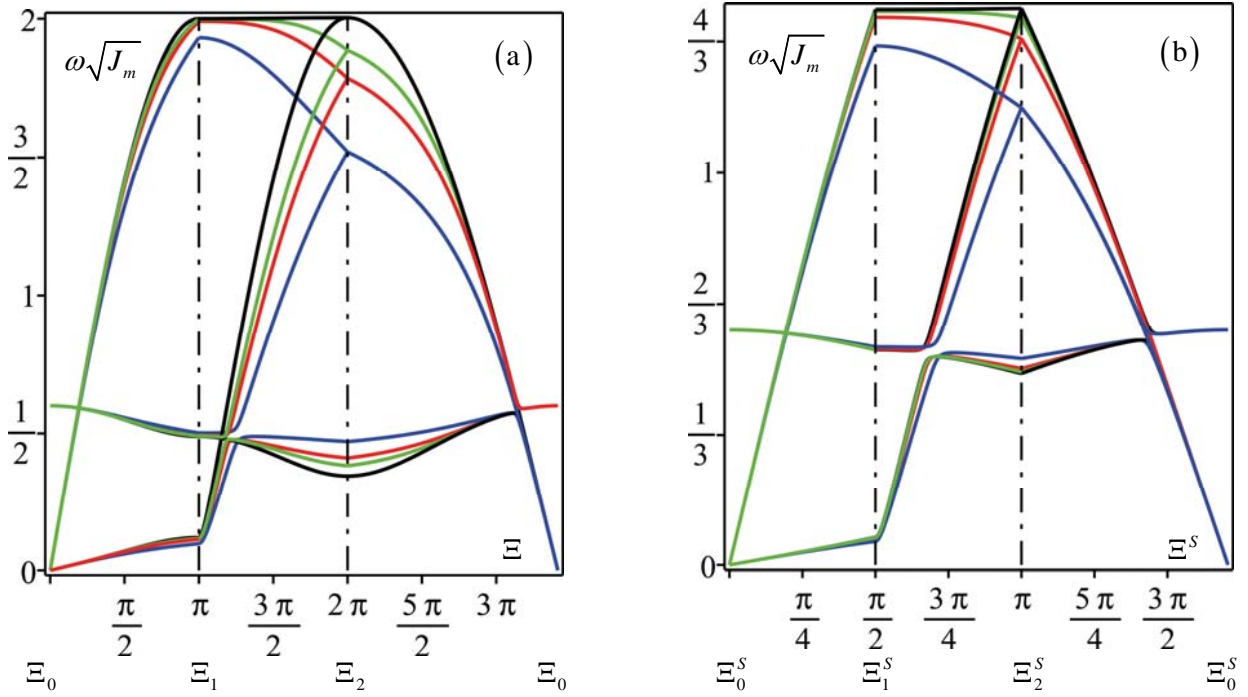

Figure 3: Frequency band structures for square beam-lattice ($r = 3/50$, $\eta = 50$). Comparison between the discrete Lagrangian model (black line) and the homogenized models obtained via 2$^{nd}$ order (blue line), 4$^{th}$ order (red line) and 6$^{th}$ order (green line) enhanced continualization. (a) Dispersive functions along the boundary of the non-dimensional irreducible first Brillouin zone; (b) Dispersive functions along the boundary of the sub-domain of the non-dimensional irreducible first Brillouin zone.

The frequency band structures for square beam-lattice characterized by $r = 3/50$ and $\eta = 50$ are shown in Figures 3.a along the closed polygonal curve $\Gamma$, both the discrete Lagrangian model (black line) and the homogenized models obtained via 2$^{nd}$ order (blue line), 4$^{th}$ order (red line) and 6$^{th}$ order (green line) enhanced continualization, respectively. It is worth to note that their dispersive functions turn out to be in good agreement with the actual corresponding one. As expected, a very good accuracy is found in the long wavelength regime, while the lesser accuracy is found for shorter wavelength regime. Moreover, the convergence of the dispersive functions obtained via enhanced continualization to the actual one is shown when increasing the continualization order. For a more complete perception of the accuracy of the homogenized models a comparison between their frequency band structures with that of the discrete Lagrangian model is given in Figure 3.b along the closed polygonal curve $\Gamma^S$ for $r = 3/50$ and $\eta = 50$.

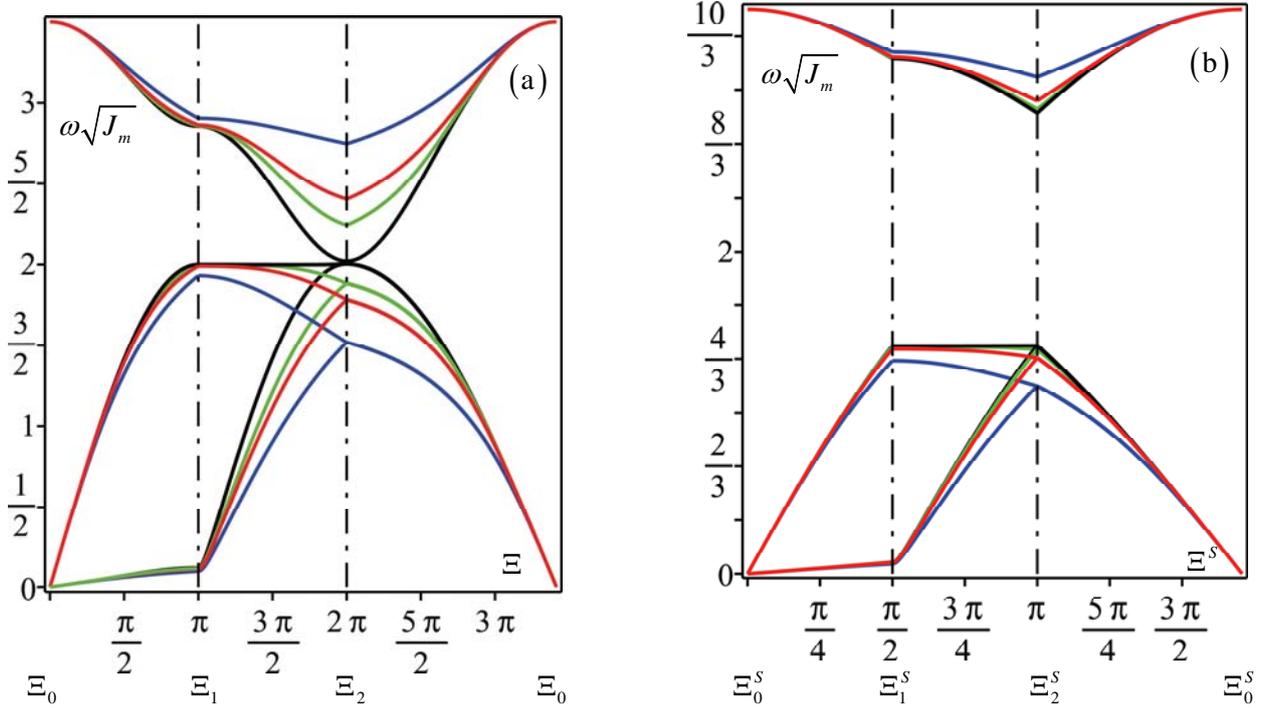

Figure 4: Frequency band structures for square beam-lattice ($r = 3/50$, $\eta = 1700$). Comparison between the discrete Lagrangian model (black line) and the homogenized models obtained via 2$^{nd}$ order (blue line), 4$^{th}$ order (red line) and 6$^{th}$ order (green line) enhanced continualization. (a) Dispersive functions along the boundary of the non-dimensional irreducible first Brillouin zone; (b) Dispersive functions along the boundary of the sub-domain of the non-dimensional irreducible first Brillouin zone.

Qualitative and quantitative analogous results in terms of the accuracy of the dispersion functions of the homogenized models obtained via enhanced continualization, and of their convergence on the frequency band structure of the discrete Lagrangian model are obtained in Figure 4 and Figure 5 for $r = 3/50$, $\eta = 1700$ and $r = 3/50$, $\eta = 3000$, respectively.

## 7. Conclusions

The main purpose of the present paper is to solve the thermodynamic inconsistencies that result when deriving equivalent micropolar models of periodic beam-lattice materials through standard continualization schemes. It is shown that this approach can lead to equivalent micropolar continuum models having non-positive defined potential energy density. This drawback affects the dynamic behavior resulting in short-wave instability phenomena and unbounded group velocity, an outcome related to the not fulfillment of the Legendre-Hadamard condition. Nevertheless, it is shown that the micropolar model thus identified provides good simulations of the frequency dispersion functions of the discrete Lagrangian system, in particular in the approximation of the optical branch of the Floquet-Bloch spectrum. Moreover, it is shown that thermodynamically consistent micropolar models derived through the extended Hamiltonian approach, i.e. equivalent to the Hill-Mandell procedure, are characterized by optical dispersion functions that do not qualitative agrees with the corresponding ones of the discrete Lagrangian system.

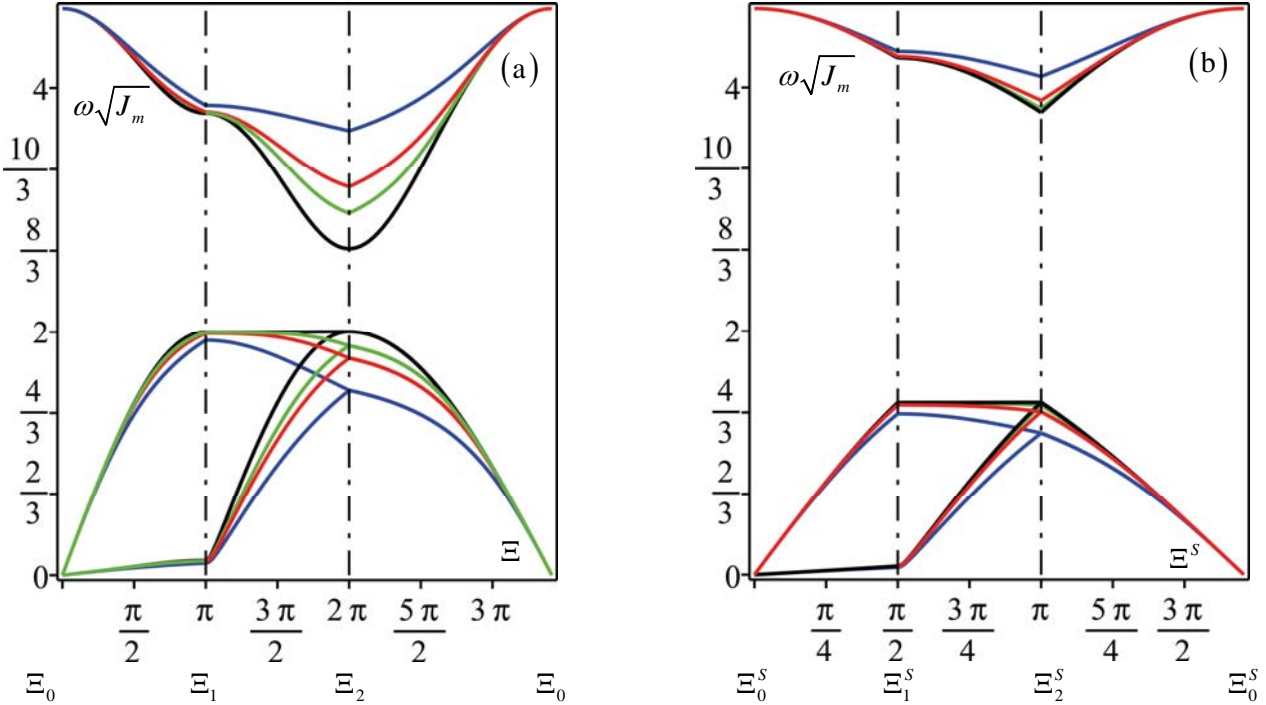

Figure 5: Frequency band structures for square beam-lattice ($r = 3/50$, $\eta = 3000$). Comparison between the discrete Lagrangian model (black line) and the homogenized models obtained via 2$^{nd}$ order (blue line), 4$^{th}$ order (red line) and 6$^{th}$ order (green line) enhanced continualization. (a) Dispersive functions along the boundary of the non-dimensional irreducible first Brillouin zone; (b) Dispersive functions along the boundary of the sub-domain of the non-dimensional irreducible first Brillouin zone.

To overcome these thermodynamic inconsistencies while preserving good simulations of the frequency band structure of the beam-lattice materials a dynamic high-frequency consistent continualization has been proposed. This enhanced continualization is based on a first order regularization approach coupled with a suitable transformation of the difference equation of motion of the discrete Lagrangian system into pseudo-differential equations. Furthermore, a formal Taylor expansion of the pseudo-differential operators allows to obtain differential field equations at various orders according to the continualization order. Thermodynamically consistent higher order micropolar continua with non-local elasticity and inertia are obtained. These models appears to be able to synthetically and accurately describe both the static and dynamic behavior of the beam-lattice materials. Finally, the convergence of the frequency band structure of the higher order micropolar models to that of the discrete Lagrangian system is shown as the continualization order increases.

**Acknowledgements**

The authors gratefully acknowledge financial support from National Group of Mathematical Physics (GNFM-INdAM), from the Compagnia San Paolo, project MINIERA no. I34I20000380007 and from University of Trento, project UNMASKED 2020.

**References**

Andrianov, I.V., Awrejcewicz J., Continuous models for 2D discrete media valid for higher-frequency domain, *Computers & structures*, **86**, 140-144, 2008

Askes H., Metrikine A.V., Higher-order continua derived from discrete media: continualisation aspects and boundary conditions, *International Journal of Solids and Structures*, **42**, 187-202, 2005.

Bacigalupo A., Gambarotta L., Simplified modelling of chiral lattice materials with local resonators, *International Journal of Solids and Structures*, **83**, 126-141, 2016.

Bacigalupo A., Gambarotta L., Wave propagation in non-centrosymmetric beam-lattices with lumped masses: discrete and micropolar modelling, *International Journal of Solids and Structures*, **118**, 128-145, 2017.

Bacigalupo, A., Gambarotta, L., Generalized micropolar continualization of 1D beam lattice, *International Journal of Mechanical Sciences*, **155**, 554-570, 2019.

Bacigalupo, A., Gambarotta, L., Identification of non-local continua for lattice-like materials, *International Journal of Engineering Science*, **159**, 103430, 2021.

Bazant Z.P., Christensen M., Analogy between micropolar continuum and grid frameworks under initial stress, *International Journal of Solids and Structures*, **8**, 327-346, 1972.

Bordiga G., Cabras L., Piccolroaz A., Bigoni D., Dynamics of prestressed elastic lattices: Homogenization, instabilities, and strain localization, *Journal of the Mechanics and Physics of Solids*, **146**, 104198, 2021.

Brillouin, L., *Wave Propagation in Periodic Structures*, 2nd ed., Dover, New York, 1953.

Carta G., Jones I.S., Movchan N.V., Movchan A.B., Wave polarization and dynamic degeneracy in a chiral elastic lattice, *Proceedings of the Royal Society A*, **475**, 20190313, 2019.

Chen J.Y., Huang Y., Ortiz M., Fracture analysis of cellular materials: a strain gradient model, *Journal of the Mechanics and Physics of Solids*, **46**, 789-828, 1998.

Chen Y., Liu X.N., Hu G.K., Sun Q.P., Zheng Q.S., Micropolar continuum modelling of bi-dimensional tetrachiral lattices, *Proc. of the Royal Society A*, **470**, 20130734, 2014.

De Domenico D., Askes H., A new multiscale dispersive gradient elasticity model with microinertia: Formulation and finite element implementation, *International Journal for Numerical Methods in Engineering*, **108,** 485-512, 2016.

De Domenico D., Askes H., Aifantis E.C., Gradient elasticity and dispersive wave propagation: Model motivation and length scale identification procedures in concrete and composite laminates, *International Journal of Solids and Structures*, **158**, 176-190, 2019.

Dos Reis F., Ganghoffer J.F., Contruction of micropolar continua from the asymptotic homogenization of beam lattices, *Computers & Structures*, **112-113**, 354-363, 2012.

Fleck N.A., Deshpande V.S., Ashby M.F., Micro-architectured materials: past, present and future, *Proceedings of the Royal Society A: Mathematical, Physical and Engineering Sciences*, **466**, 2495-2516, 2010.

Gad A.I., Gao X.L., Li K., A Strain Energy-Based Homogenization Method for 2-D and 3-D Cellular Materials Using the Micropolar Elasticity Theory, Composite Structures, **20**, 113594, 2021.

Gibson, L. J., and Ashby, M. F., *Cellular Solids: Structure and Properties*, Cambridge University Press, Cambridge, 1997.

Gómez-Silva F., Fernández-Sáez J., Zaera R., Nonstandard continualization of 1D lattice with next-nearest interactions. Low order ODEs and enhanced prediction of the dispersive behavior. *Mechanics of Advanced Materials and Structures,* Sep 8:1-0, 2020.

Gómez-Silva F., Zaera R., Analysis of low order non-standard continualization methods for enhanced prediction of the dispersive behaviour of a beam lattice, *International Journal of Mechanical Sciences,* **196**, 06296, 2021.

Gonnella S., Ruzzene M., Homogenization and equivalent in-plane properties of two-dimensional periodic lattices, *International Journal of Solids and Structures*, **45**, 2897–2915, 2008.

Jordan C., *Calculus of finite differences*, Vol. 33. American Mathematical Soc., 1965.

Kamotski I.V., Smyshlyaev V.P., Bandgaps in two-dimensional high-contrast periodic elastic beam lattice materials, *Journal of the Mechanics and Physics of Solids*, **123**, 292-304, 2019.

Kelley W.G., Peterson A.C., *Difference equations. An introduction with applications*, Academic Press, 2001.

Krödel S., Delpero T., Bergamini A., Ermanni P., Kochman D.M., 3D Auxetic Microlattices with Independently Controllable Acoustic Band Gaps and Quasi-Static Elastic Moduli, *Advanced Engineering Materials*, **16**, 357-363, 2014.

Kumar R.S., McDowell D.L., Generalized continuum modelling of 2-D periodic cellular solids, *International Journal of Solids and Structures*, **41**, 7399-7422, 2004.

Lombardo M., Askes H., Higher-order gradient continuum modelling of periodic lattice materials, *Computational Material Science*, **52**, 204-208, 2012.

Lu Z., Wang Q., Li X., Yang Z., Elastic properties of two novel auxetic 3D cellular structures, *International Journal of Solids and Structures*, **124**, 46-56, 2017.

Martisson P.G., Movchan A.B., Vibration of lattice structures and phononic band gaps, *Quarterly Journal of Mechanics and Applied Mathematics*, **56**, 45–64, 2003.

Ostoja-Starzewski M., Lattice models in micromechanics, *Applied Mechanics Review*, **55**, 35-60, 2002.

Phani A.S., Woodhouse J., Fleck N.A., Wave propagation in two-dimensional periodic lattices, *J. Acoustical Society of America*, **119**, 1995-2005, 2006.

Piccolroaz A., Movchan A.B., Cabras L., Rotational inertia interface in a dynamic lattice of flexural beams, *International Journal of Solids and Structures*, **112**, 43-53, 2017.

Prall D., Lakes R.S., Properties of chiral honeycomb with a Poisson ratio of -1, *Int. J. Mechanical Sciences*, **39**, 305-314, 1997.

Seiler P.E., Li K., Deshpande V.S., Fleck N.A., The influence of strut waviness on the tensile response of lattice materials, *Journal of Applied Mechanics*, **88**, 031011, 2021.

Suiker A.S.J., Metrikine A.V., de Borst R., Comparison of wave propagation characteristics of the Cosserat continuum model and corresponding discrete lattice models, *International Journal of Solids and Structures*, **38**, 1563-1583, 2001.

Vasiliev A.A., Dmitriev S.V., Miroshnichenko A.E., Multi-field approach in mechanics of structural solids, *International Journal of Solids and Structures*, **47**, 510-525, 2010.

Vasiliev A.A., Miroshnichenko A.E., Ruzzene M., Multifield model for Cosserat media, *J. of Mechanics of Materials and Structures*, **3**, 1365-1382, 2008.